\documentclass[aps,prl,reprint,superscriptaddress,groupedaddress,letterpaper,citeautoscript]{revtex4-1}
\usepackage[british]{babel}
\usepackage{graphicx}
\usepackage[pdftex                     % using pdftex
           ,pagebackref=false          % put link back to original page ?
           ,colorlinks=false           % color link numbers
           ,hidelinks
           ]{hyperref}
\hypersetup{linkcolor=blue,            % figures, equations, refs
            citecolor=red,             % citations
            urlcolor=black}            % URLs

\usepackage{amsmath}
\usepackage{mathtools}
\usepackage{eufrak}
\usepackage{times}

\newcommand*{\vn}{\boldsymbol}

\begin{document}

\begin{abstract}
We predict from first principles an entirely topological orbital magnetization in the noncoplanar bulk antiferromagnet $\gamma$-FeMn originating in the nontrivial topology of the underlying spin structure, without any reference to spin-orbit interaction. Studying the influence of strain, composition ratio, and spin texture on the topological orbital magnetization and the accompanying topological Hall effect, we promote the scalar spin chirality as key mechanism lifting the orbital degeneracy. The system is thus a prototypical {\it topological orbital ferromagnet}, the macroscopic orbital magnetization of which is prominent even without spin-orbit coupling. One of the remarkable features of $\gamma$-FeMn is the possibility for pronounced orbital magnetostriction mediated by the complex spin topology in real space.
\end{abstract}

\setcounter{secnumdepth}{3}
 \title{
Prototypical topological orbital ferromagnet \texorpdfstring{$\gamma$}{gamma}-FeMn}
 \author{Jan-Philipp Hanke}
 \email{j.hanke@fz-juelich.de}
 \author{Frank Freimuth}
 \author{Stefan Bl\"ugel}
 \author{Yuriy Mokrousov}
 \affiliation{Peter Gr\"unberg Institut and Institute for Advanced Simulation,\\Forschungszentrum J\"ulich and JARA, 52425 J\"ulich, Germany}
 \maketitle

\section{Introduction}
Spin and orbital degrees of freedom of electrons give rise to two fundamental contributions to the magnetization in materials, both of which are usually distinguished by means of x-ray magnetic circular dichroism~\cite{Thole1992,Carra1993,Chen1995}. While firm knowledge of spin magnetism has been acquired due to extensive research in this area over the past several decades, exploration of the concept of the orbital magnetization (OM) in condensed matter is still at a rather premature stage. Even an accurate theoretical description of orbital magnetism has been missing until the recent advent of a rigorous Berry phase theory~\cite{Thonhauser2005,Ceresoli2006,Xiao2005,Shi2007}. Since the OM affects a phlethora of elementary properties like spin-dependent transport~\cite{Xiao2006,Murakami2006,Xiao2007,Wang2007}, orbital magnetoelectric coupling~\cite{Malashevich2010,Essin2009,Essin2010}, and Dzyaloshinskii-Moriya interaction~\cite{Moriya1960}, a deeper understanding of orbital magnetism in solids is in general of outstanding relevance.

Spontaneous orbital magnetism in ferromagnets is conventionally explained as a key manifestation of the spin-orbit interaction lifting (partially) the orbital moment quenching. While such an interpretation applies to most condensed-matter systems, it fails to describe orbital magnetism in crystals exhibiting a finite {\it topological} OM (TOM) prominent even in the absence of spin-orbit coupling. In these systems an emergent magnetic field rooting in the noncoplanarity of neighbouring spins replaces the spin-orbit interaction as the main mechanism lifting the orbital degeneracy by coupling to the orbital degrees of freedom~\cite{Shindou2001,Hoffmann2015,Hanke2016}. The latter noncoplanarity between three neighbouring spins is usually quantified by the so-called scalar spin chirality $\kappa=\vn S_i \cdot (\vn S_j\times \vn S_k)$, which plays also a fundamental role in the physics of skyrmions~\cite{Bruno2004,Neubauer2009,Kanazawa2011,Huang2012,Franz2014,Gayles2015,Dias2016}. The nonvanishing scalar spin chirality further replaces the spin-orbit interaction in giving rise to the anomalous Hall effect, also referred to as topological Hall effect in this context, which is closely related to the TOM both from microscopic and symmetry 
considerations~\cite{Shindou2001,Taguchi2001}. 
\begin{figure}
\centering
\scalebox{0.15}{\includegraphics{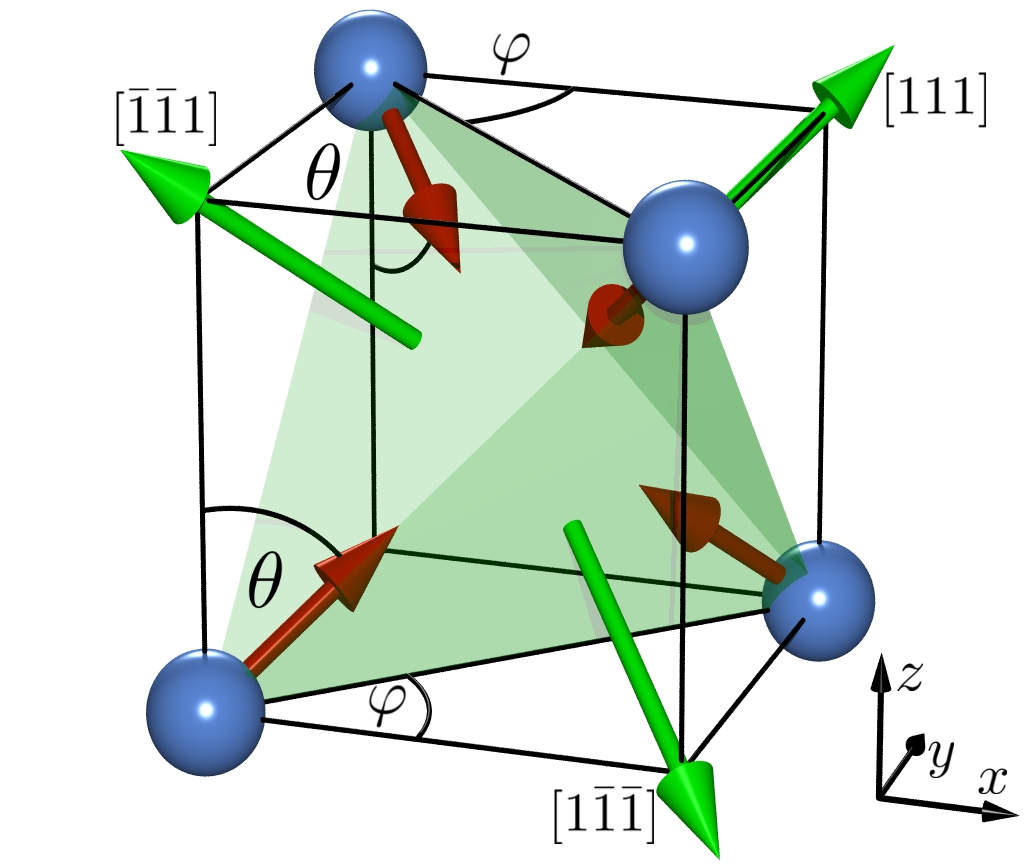}}
\caption{{\bf Unit cell of face-centered-cubic $\gamma$-FeMn}. Red arrows indicate the noncoplanar $3\vn Q$ spin texture characterized by the polar angle $\theta=54.74^\circ$ and the azimuthal angle $\varphi=45^\circ$. Green triangles spanned by neighbouring spins highlight equivalent planes of the undistorted lattice, and green arrows orthogonal to corresponding planes mark the directions $[\bar 1\bar 11]$, $[1\bar 1 \bar 1]$, and $[111]$, respectively.}
\label{fig:structure}
\end{figure}

Revealing a diverse spectrum of complex noncollinear spin textures in real space, antiferromagnetic materials such as Mn$_5$S$_3$~\cite{Brown1992,Suergers2014,Suergers2016}, Nd~\cite{Bak1979,Forgan1982,McEwen1986,Forgan1989}, Mn$_3$GaN~\cite{Bertaut1968,Gomonay2015}, Mn$_3$Ir~\cite{Chen2014}, and Mn$_3$Ge~\cite{Kuebler2014} provide an intriguing and rich playground to study unconventional magnetic properties and transport phenomena. In particular, disordered $\gamma$-Fe$_x$Mn$_{1-x}$ alloys are suggested by neutron diffraction measurements~\cite{Kouvel1963,Endoh1971} and first-principles calculations~\cite{Kuebler1988,Schulthess1999,Sakuma2000} to exhibit the so-called $3\vn Q$ structure, which is a noncoplanar spin configuration forming as a result of a linear combination of spin-spirals with three distinct wave vectors  $\vn Q$ (cf. Fig.~\ref{fig:structure}) \footnote{In the case of $\gamma$-FeMn with simple cubic unit cell containing four atoms as shown in Fig.~\ref{fig:structure}, the corresponding wave vectors are $\vn Q_1=(2\pi/a,0,0)$, $\vn Q_2=(0,2\pi/a,0)$, and $\vn Q_3=(0,0,2\pi/a)$, where $a$ is the lattice constant.}. In the latter case of such a $3\vn Q$ state the total spin magnetization integrated over the unit cell of the crystal vanishes exactly. The $3\vn Q$ noncollinear spin structure of disordered $\gamma$-Fe$_x$Mn$_{1-x}$ alloys thus renders these systems ideal candidates for investigating topological contributions to the OM and the accompanying anomalous Hall conductivity (AHC), as well as for estimating the efficiency of the scalar spin chirality as alternative degeneracy-breaking mechanism.

\begin{figure*}
\centering
\includegraphics{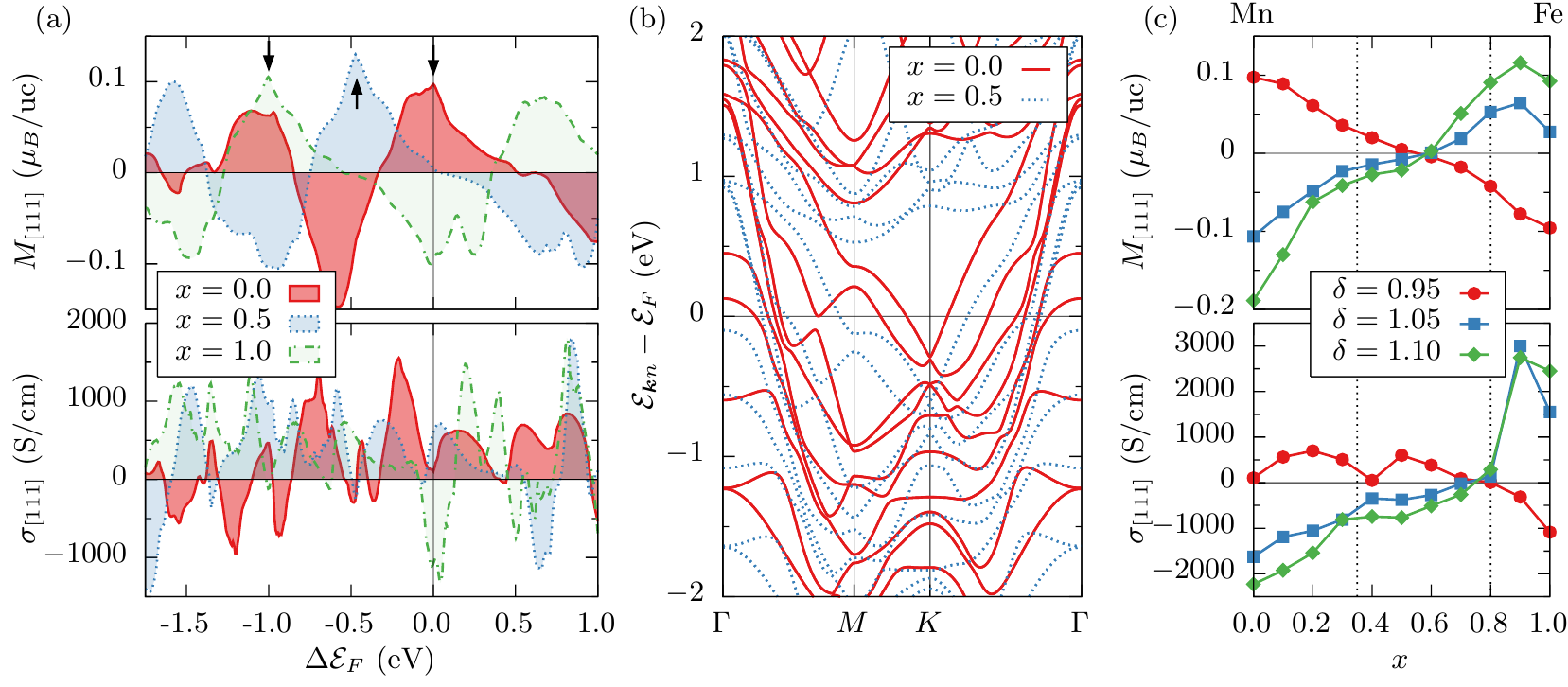}
\caption{{\bf Influence of strain and composition ratio}. ({\bf a}) The only nonvanishing component of orbital magnetization (OM) and anomalous Hall conductivity (AHC) at the strain $\delta=0.95$ as a function of the Fermi energy for the $3\vn Q$ spin structure and several concentrations $x$
of the $\gamma$-Fe$_x$Mn$_{1-x}$ alloy. Arrows mark characteristic features, which shift towards lower energies with increasing $x$. ({\bf b}) The corresponding electronic band structures for $x=0.0$ and $x=0.5$ exhibit strong similarities due to the close nature of Fe and Mn. ({\bf c}) The only nonvanishing component of OM and AHC at the actual Fermi energy as a function of the concentration $x$ for the $3\vn Q$ spin structure and various strains $\delta$. Dotted vertical lines bound the experimentally confirmed region of the $3\vn Q$ state. In all cases ({\bf a}-{\bf c}) spin-orbit coupling was not included in the calculations, and the unit cell (uc) is the magnetic unit cell of four atoms.}
\label{fig:energy}
\end{figure*}
Here, based on first-principles electronic-structure calculations we evaluate topological orbital magnetism and anomalous Hall effect present in the noncoplanar antiferromagnet $\gamma$-FeMn even in absence of spin-orbit coupling. We ascribe the manifestation of these phenomena to the nontrivial topology of the underlying $3\vn Q$ spin structure. By studying the dependencies of both OM and AHC on (i) strain, (ii) composition ratio, and (iii) real space distribution of the spin texture, we identify the scalar spin chirality as driving mechanism which lifts the orbital degeneracy. Thus, disordered $\gamma$-FeMn is a prototypical {\it topological orbital ferromagnet} for which the macroscopic magnetization is completely dominated by orbital magnetism prominent even without spin-orbit interaction.

\section{Results}
Although the noncoplanar spin texture of the $3\vn Q$ state gives rise to a finite scalar spin chirality between any three neighbouring spins (cf. Fig.~\ref{fig:structure}), the sum of these chiralities over the magnetic unit cell containing four atoms vanishes in the case of the face-centered-cubic (fcc) lattice of $\gamma$-FeMn. Thus, to investigate possible topological contributions arising from the nontrivial spin texture in real space, we need to break the symmetry between the equivalent cross sections $(111)$, $(\bar{1}\bar{1}1)$, $(\bar{1}1\bar{1})$, and $(1\bar{1}\bar{1})$ of the fcc lattice (see Fig.~\ref{fig:structure}), for example, by applying strain along the $[111]$ direction~\cite{Shindou2001}. As a consequence of this distortion, the chiralities of the individual cross sections do not cancel out such that we can expect the scalar spin chirality to manifest in the $[111]$ component of both OM and AHC. To characterize the deformation of the crystal structure, we introduce the ratio $\delta=d^\prime/d$ where $d$ and $d^\prime$ refer to the distance between adjacent $(111)$ planes in the undistorted and distorted case, respectively.

Applying a moderate strain along the $[111]$ direction characterized by the parameter $\delta=0.95$, we present in Fig.~\ref{fig:energy}a the resulting computed OM and AHC of selected antiferromagnetic $\gamma$-Fe$_x$Mn$_{1-x}$ alloys as a function of the position of the Fermi level and in absence of spin-orbit coupling. When we change the alloy composition from pure Mn over Fe$_{0.5}$Mn$_{0.5}$ to pure Fe, we observe that the energy dependence of the OM is qualitatively not altered except for a shift towards lower energies (see~e.g.~peaks marked by arrows in Fig.~\ref{fig:energy}a). Since the electronic configurations of Fe and Mn atoms are close to each other, the electronic structures of the corresponding $\gamma$-Fe$_x$Mn$_{1-x}$ alloys are quite similar. This can be directly seen in the band structure of the corresponding alloys, which reveal a clear resemblance apart from a global energy shift, Fig.~\ref{fig:energy}b. Remarkably, although OM and anomalous Hall effect are fundamentally related to each other, the energy dependencies of OM and AHC presented in Fig.~\ref{fig:energy}a are not at all correlated. In contrast to the OM, the energy dependence of the AHC displays a more rich and complex oscillatory behavior underlining its strong sensitivity to the electronic structure. Moreover, the energy dependencies of the AHC for different alloy concentrations are not related by a simple shift in energy. We verified that the values of both OM and AHC, which we display in Fig.~\ref{fig:energy}a, are hardly affected by including spin-orbit coupling in our calculations.

We further present in Fig.~\ref{fig:energy}c the behavior of OM and AHC at the actual Fermi energy of $\gamma$-Fe$_x$Mn$_{1-x}$, i.e., $\Delta\mathcal E_F=0$, upon varying the concentration $x$. A moderate contraction ($\delta=0.95$) of the pure Mn crystal generates an OM of $0.1\mu_B$ per unit cell of four atoms, which monotonically decreases with increasing Fe concentration, leading to a sign change for $x\approx 0.5-0.6$ and ultimately to an OM of about $-0.1\mu_B$ per unit cell for pure Fe. Remarkably, this overall dependence on the alloy composition is reversed if we expand the lattice ($\delta>1$) along the $[111]$ direction instead of contracting it. Within the experimentally confirmed region of the $3\vn Q$ state in $\gamma$-FeMn, the OM ranges from $-0.05$ to $0.1\mu_B$ per unit cell depending on the strain. In general, we find that applying larger strain leads to an enhanced magnitude of both OM and AHC since the symmetry-breaking between the formerly equivalent faces of the tetrahedron spanned by the four spins in the unit cell becomes more pronounced in this case. The concentration dependence of the AHC qualitatively follows the trend of the OM, in particular for $\delta>1$. We point out that the magnitude of the AHC in these alloys reaches gigantic values, especially in the vicinity of pure alloys, constituting as much as $1000$\,S/cm in the experimentally confirmed region of the $3\vn Q$ state in $\gamma$-Fe$_x$Mn$_{1-x}$ (see dashed lines in Fig.~\ref{fig:energy}c). Such strong anomalous Hall effect in a spin-compensated material is remarkable, and we motivate experimental studies aimed at its detection. We checked that the influence of spin-orbit coupling on the values of OM and AHC, as presented in Fig.~\ref{fig:energy}c, is negligible. This promotes the noncoplanarity of the underlying spin texture as driving mechanism for these phenomena in the bulk antiferromagnet $\gamma$-FeMn.

\begin{figure}
\centering
\includegraphics{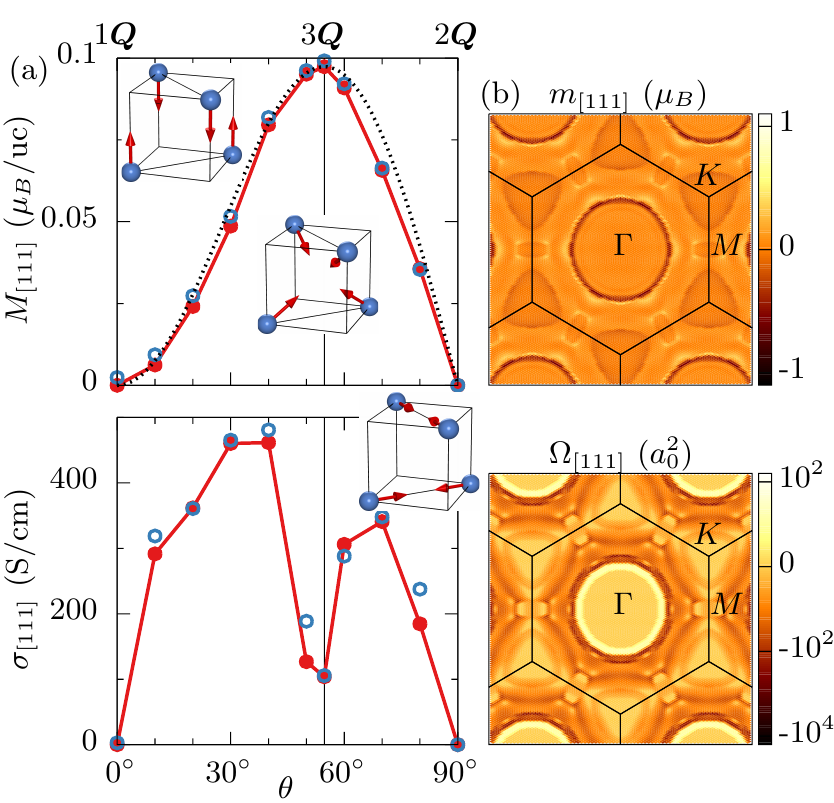}
\caption{{\bf Influence of spin texture.} ({\bf a}) The $[111]$ component of orbital magnetization and anomalous Hall conductivity as a function of the polar angle $\theta$, which characterizes the antiferromagnetic spin texture, for the strain $\delta=0.95$ and the concentration $x=0.0$. Spin-orbit coupling was additionally taken into account for the open blue data points. The dotted black line denotes the angular dependence of the spin chirality, Eq.~\eqref{eq:chirality_ideal_contribution}, and the insets present the multiple-$\vn Q$ structures at $\theta=0^\circ$, $54.74^\circ$, and $90^\circ$. The unit cell (uc) is the magnetic unit cell containing four atoms. ({\bf b}) Projected reciprocal-space distribution of the orbital moment $m_{[111]}$ and the Berry curvature $\Omega_{[111]}$ for the $3\vn Q$ state in absence of spin-orbit coupling. Note the logarithmic color scale in case of the Berry curvature.}
\label{fig:angle}
\end{figure}

We show in Fig.~\ref{fig:angle}b the reciprocal-space distribution of the orbital moment and the Berry curvature of the occupied states for the $3\vn Q$ state of the pure Mn system with $\delta=0.95$, projected onto the $(111)$ plane. Both quantities reveal pronounced sharp features associated with the Fermi surface lines (cf. Fig.~\ref{fig:energy}b), and reflect the six-fold rotational symmetry associated with the $(111)$ plane of the distorted lattice. While singular large negative contributions to the OM forming a circle around the $\Gamma$ point are due to the group of parabolic bands which cross the Fermi level close to $\Gamma$ along the paths $\Gamma M$ and $\Gamma K$ in Fig.~\ref{fig:energy}b, the overall positive OM in this case is determined by large areas of the Brillouin zone providing small positive contributions. In analogy to the case of the energy dependence discussed before, the Berry curvature generally reveals a richer and more complex distribution in reciprocal space as compared to the orbital moment.

Finally, we explicitly scrutinize the effect of the antiferromagnetic spin distribution in real space on orbital magnetism and anomalous Hall effect. We do this by varying the polar angle $\theta$ which defines the details of the spin texture (cf. Fig.~\ref{fig:structure}), while keeping the other characteristic angle $\varphi$ fixed to the value of $45^\circ$, as in the calculations reported above. Tuning the parameter $\theta$ effectively allows us to control the value of the scalar spin chirality $\kappa$ between three neighbouring spins according to the following relation:
\begin{equation}
 \kappa(\theta) \propto \cos\theta\sin^2\theta \, .
 \label{eq:chirality_ideal_contribution}
\end{equation}
Apparently, $\kappa(\theta)$ is zero for the cases of $\theta=0^\circ$ (also known as the $1\vn Q$ state) and
$\theta=90^\circ$ (known as the $2\vn Q$ state), and it becomes maximal for $\theta=54.74^\circ$ in the $3\vn Q$ structure as indicated by the thin dotted line in Fig.~\ref{fig:angle}a. In the latter figure also the dependence of the $[111]$ component of both OM and AHC on the polar angle $\theta$ for the pure Mn system with $\delta=0.95$ is shown. Reaching its maximum of about $0.1\mu_B$ per unit cell for the $3\vn Q$ structure, the $\theta$-dependence of the OM falls almost perfectly onto the $\kappa(\theta)$ curve. The small deviations between the two quantities are the
consequence of changes in the fine details of the electronic structure of the system as $\theta$ is varied. The nice agreement between the OM and $\kappa$ in $\gamma$-FeMn is a clear indication that in this class of compounds the OM is directly proportional to the emergent magnetic field as given
by the scalar spin chirality, with the constant of proportionality playing the role of a {\it topological orbital susceptibility}. On the other hand, the dependence of the AHC on the polar angle $\theta$ is more complex and cannot be directly related to the simple angular dependence of $\kappa(\theta)$ but originates rather in higher-order terms. As a result, in contrast to the OM, the value of the AHC exhibits a local minimum for the $3\vn Q$ spin texture. To prove that the origin of both OM and AHC lies purely in the real-space distribution of spins, we perform calculations including the effect of spin-orbit coupling. We find that the OM values hardly change upon taking the spin-orbit interaction into account, and likewise the corresponding changes in the AHC driven by the modification of the electronic structure due to spin-orbit interaction are insignificant, see Fig.~\ref{fig:angle}a.

\section{Discussion}
Our results clearly demonstrate that the noncoplanar bulk antiferromagnet $\gamma$-FeMn is a prototypical {\it topological orbital ferromagnet} (TOF). The emergent magnetic field associated with the scalar spin chirality of the $3\vn Q$ state completely replaces the spin-orbit coupling as lifting mechanism of the orbital moment quenching. In TOFs of the type discussed here, nonzero charge and orbital currents are the consequence of the spin chirality in the distorted crystal leading to purely topological contributions to orbital magnetism and anomalous Hall effect. We further expect noncoplanar magnets to exhibit analogously a {\it topological spin Hall effect} (see also Ref.~\cite{Kang2016}) stemming solely from the nontrivial topology of the spin texture, without any reference to spin-orbit coupling.

In a wider context, as compared to the spin of electrons, the orbital degrees of freedom offer higher flexibility regarding their internal structure and the size of orbital moments, rendering them versatile operational building blocks in the field of {\it orbitronics}. In this respect, TOFs as a new class of materials occupy a special place since their nontrivial orbital magnetism is a direct consequence of complex spin arrangement. This means that the properties of TOFs can be directly tuned by altering the latter spin distribution, e.g., via electric-field-induced spin torques~\cite{Wadley2016} or by modifying the strength of the spin-spin interactions. Our calculations also clearly indicate that both magnitude and sign of the TOM in TOFs can be controlled efficiently by means of proper electronic-structure engineering through application of strain and variation of the concentration in alloys. The magnitude of $0.1\mu_B$ per unit cell in the studied $\gamma$-FeMn can be enhanced even further by considering thin films with noncollinear spin textures~\cite{Hanke2016}.

While the topological Hall effect relates to higher-order terms in $\kappa$, the connection between the finite TOM and the scalar spin chirality as driving force is particularly intimate as apparent from Fig.~\ref{fig:angle}a. Therefore, we speculate that an inverse effect should be observable in terms of an increased magnetostructural coupling in noncoplanar antiferromagnets~\cite{Peng2006,He2009}. That is, we claim that the systems of the type discussed here will experience a pronounced {\it topological orbital magnetostriction} due to the interaction of the TOM with an applied external magnetic field, leading to changes in the crystallographic structure as a result of maximizing the energy gain due to this interaction. In particular, owing to the opposite sign of the TOM with 
respect to tensile or compressive strain along $[111]$ direction in $\gamma$-FeMn, it would be possible to expand or compress the crystal along the [111] axis by changing the orientation of an applied magnetic field along this axis.

\section{Methods}
Using the full-potential linearized augmented plane-wave code \texttt{FLEUR}~\cite{fleur}, we performed self-consistent density functional theory calculations of the electronic structure of disordered $\gamma$-FeMn alloys within the virtual crystal approximation by adapting the nuclear numbers under conservation of charge neutrality. Exchange and correlation effects were treated in the generalized-gradient approximation of the PBE functional~\cite{PBE}. We have chosen a muffin-tin radius of $2.29a_0$ and the lattice constant of the undistorted fcc lattice was $6.86a_0$. When we applied strain to expand or contract the lattice along the $[111]$ direction, according deformations of the transverse crystal directions were taken into account employing the Poisson's ratio $\nu=0.27$. The plane-wave cutoff was $3.8a_0^{-1}$ and the Brillouin zone was sampled using an equidistant $12\times 12 \times 12$-mesh of $\vn k$ points. The calculated local spin moments amount to about $2\mu_B$ depending on the applied strain and the imposed spin structure.

To unambiguously compute the OM, we applied a recently developed Berry phase theory~\cite{Thonhauser2005,Ceresoli2006,Xiao2005,Shi2007} expressing the OM as a genuine bulk property of the ground-state wave functions:
\begin{equation}
 \vn M =\frac{e}{2\hbar} \text{Im} \int[\mathrm dk]\, \langle \partial_{\vn k} u_{\vn kn}| \times\left(H_{\vn k} + \mathcal E_{\vn kn} - 2 \mathcal E_F\right)| \partial_{\vn k} u_{\vn kn}\rangle \, ,
 \label{eq:orbmom}
\end{equation}
where $\vn k$ is the crystal momentum, $[\mathrm d k]$ stands for $\sum_n^{\text{occ}}\mathrm d\vn k /(2\pi)^3$ with the summation restricted to all occupied bands $n$ below the Fermi energy $\mathcal E_F$, $|u_{\vn kn}\rangle$ is an eigenstate of the lattice-periodic Hamiltonian $H_{\vn k}= \mathrm e^{-i\vn k \cdot \vn r} H \mathrm e^{i \vn k \cdot \vn r}$ to the band energy $\mathcal E_{\vn kn}$, and $e>0$ is the elementary positive charge. Within this framework, the AHC can be expressed as
\begin{equation}
 \vn \sigma = -\frac{e^2}{\hbar} \text{Im} \int[\mathrm dk]\, \langle \partial_{\vn k} u_{\vn kn}| \times| \partial_{\vn k} u_{\vn kn}\rangle \, .
 \label{eq:cond}
\end{equation}
Both orbital moment $\vn m$ and Berry curvature $\vn \Omega$ in reciprocal space as displayed in Fig.~\ref{fig:angle}b were obtained from the relations $\vn M = \int \vn m(\vn k) [\mathrm dk]$ and $\vn \sigma = e^2/\hbar \int \vn \Omega(\vn k) [\mathrm dk]$, respectively.

The Brillouin zone integration in Eqs.~\eqref{eq:orbmom} and~\eqref{eq:cond} can be performed efficiently through Wannier interpolation~\cite{Wang2006,Yates2007,Lopez2012}. Based on an $8\times 8\times 8$ $\vn k$-mesh, we constructed $72$ maximally-localized Wannier functions out of $102$ energy bands~\cite{Mostofi2014,Freimuth2008} with the frozen window extending up to about $5\,\text{eV}$ above the Fermi energy. Finally, convergence of OM and AHC was achieved using interpolation meshes of $128\times 128\times 128$ and $256\times 256 \times 256$ $\vn k$-points, respectively.

\section*{Acknowledgements}
\noindent
We gratefully acknowledge computing time on the supercomputers JUQUEEN and JURECA at
J\"ulich Supercomputing Center as well as at the JARA-HPC cluster of RWTH Aachen, and funding under SPP 1538 of Deutsche Forschungsgemeinschaft (DFG) as well as funding from the European Union’s Horizon 2020 research and innovation programme under grant agreement number 665095 (FET-Open project MAGicSky).

%\section*{Author contributions}
%\noindent
%J.-P.H. performed the first-principles calculations and analysed the results. J.-P.H. and Y.M. wrote the manuscript. All authors discussed the results and reviewed the manuscript.

%\section*{Additional Information}
%\noindent
%{\bf Competing financial interests:} The authors declare no competing financial interests.

\bibliography{my_bibliography}

\end{document}